\documentclass[preprint,showpacs,preprintnumbers,amsmath,amssymb,endfloats]{revtex4}

\usepackage{graphicx}

\newcommand{\nn}{\noindent}
\newcommand{\bes}{\begin{eqnarray}}
\newcommand{\ees}{\end{eqnarray}}

\begin{document}

\thispagestyle{empty}
\title{
Improved  tests of extra-dimensional physics and thermal quantum
field theory from new Casimir force measurements
}
\author{R.~S.~Decca,${}^{1}$ E.~Fischbach,${}^{2}$\footnote{Corresponding
author.}
   G.~L.~Klimchitskaya,${}^{3}$\footnote {On leave from
North-West Technical University, St.\ Petersburg, Russia}
D.~E.~Krause,${}^{4,2}$ D.~L\'{o}pez,${}^{5}$ and
V.~M.~Mostepanenko${}^{3}$\footnote{On leave from Noncommercial
Partnership ``Scientific Instruments'', Moscow, Russia}
}

\affiliation{${}^{1}$Department of Physics, Indiana University-Purdue
University Indianapolis, Indianapolis, Indiana 46202, USA \\
${}^{2}$Department of Physics, Purdue University, West Lafayette, Indiana
47907, USA \\
${}^{3}$Departamento de F\'{\i}sica, Universidade Federal da Para\'{\i}ba,
C.P.5008, CEP 58059--970, Jo\~{a}o Pessoa, Pb-Brazil \\
${}^{4}$Physics Department, Wabash College, Crawfordsville, Indiana 47933,
USA \\
${}^{5}$Bell Laboratories, Lucent Technologies, Murray Hill,
New Jersey 07974, USA
}

\begin{abstract}
We report new constraints on extra-dimensional models and other physics
beyond the Standard Model based on measurements of the Casimir force
between two dissimilar metals  for separations in the range
0.2--1.2\,$\mu$m.  The Casimir force between an Au-coated
sphere and a Cu-coated plate of a microelectromechanical torsional
oscillator was measured statically with an absolute error of 0.3\,pN.  In
addition,  the Casimir pressure between two
parallel plates was determined dynamically with an absolute error of
$\approx 0.6$\,mPa.
Within the limits of experimental and theoretical errors, the
results are in agreement with a theory that takes into account the finite
conductivity and roughness of the two metals. The level of agreement between
experiment and theory was then used to set limits on the predictions of
extra-dimensional physics and thermal quantum field theory. It is shown that
two theoretical approaches to the thermal Casimir force which predict effects
linear in temperture are ruled out by these experiments.
Finally, constraints on Yukawa corrections to Newton's
law of gravity are strengthened by more than an order of
magnitude in the range  56\,nm to 330\,nm.
\end{abstract}


\maketitle

\section{Introduction}

Many extensions of the Standard Model,
including supergravity and string theory, exploit the Kaluza-Klein
idea that the true dimensionality of space-time is $N=4+n$, where
the additional $n$ spatial dimensions are compactified at some small
length scale. For a long time it was generally believed that the
compactification scale was on the order of the Planck length
$l_{Pl}=\sqrt{G}\sim 10^{-33}\,$cm, where $G$ is the Newtonian
gravitational constant, and units are chosen so that
$\hbar=c=1$.  The corresponding energy scale
$M_{Pl}=1/\sqrt{G}\sim 10^{19}\,$GeV is so high that direct
experimental observation of the effects of extra dimensions would
seem impossible at any time in the foreseeable future.

The situation changed dramatically with the proposal of models  for
which the compactification energy may be as low as the
extra-dimensional Planck energy scale,
$M_{Pl}^{(N)}=1/G_{N}^{1/(2+n)}$, which is assumed to be of order
1\,TeV \cite{1,2}.   (Here $G_{N}$ is the fundamental gravitational
constant in the extended $N$-dimensional space-time).  Note that this
proposal eliminates the hierarchy problem since the characteristic
energy scales of gravitational and gauge interactions coincide.
In order to be consistent with observations, the usual gauge fields of
the Standard Model are presumed to exist on  4-dimensional branes
whereas gravity alone propagates into the
$N$-dimensional bulk.

Constraints on these new lower energy scale
compactification models can be obtained by investigating their
predictions in accelerator experiments \cite{3,4,5}, astrophysics
\cite{6,7,8}, and cosmology \cite{9,10,11}.  More
model-independent limits, however,
can be obtained from tests of Newtonian
gravity.   In extra-dimensional models with large compact extra
dimensions \cite{1}, the Newtonian
gravitational potential acting between two point masses acquires a
Yukawa correction for separations much larger than the
compactification scale
\cite{12,13}, while for  models with non-compact but warped extra
dimensions,  the corrections are power-laws \cite{2}.
For two interacting macroscopic bodies, either of these corrections
would give rise to a  new (so called ``fifth'') force coexisting with the
usual  Newtonian gravitational force and other conventional Standard Model
interactions, such as Casimir and van der Waals forces.   In addition,  many
extensions of the Standard Model that do not involve extra dimensions also
predict the existence of new Yukawa or power-law forces.

While gravity experiments at ranges $\gtrsim 10^{-3}$~m have found no
convincing evidence of new forces or extra-dimensions, tests of
Newtonian gravity over shorter separations were lacking until
recently.   During the past few years a number of
sub-millimeter gravity experiments were  performed and stronger
constraints on Yukawa corrections to  Newtonian
gravity for ranges $\sim 10^{-4}\,$m have been obtained
\cite{14,15,16,17}.  For significantly smaller separations, however,
gravity loses its role as the
dominant force acting between non-magnetic, electrically neutral
interacting bodies.
For these smaller separations,  limits on new forces and
extra-dimensions from force measurements must be extracted from the
Casimir and van der Waals forces \cite{18} which increase rapidly
as the separation decreases.

Improvements in the precision of Casimir force measurements
coincided with the development of modern extra-dimensional
theories.  To date a number of  Casimir force experiments have been
performed using different
techniques \cite{19,20,21,22,23,24,25,26,27}, and a precision of 2-3\%
(at the 95\% confidence level) of the
measured force at the shortest separation distances has been achieved
(see Ref.~\cite{27a}).  We note that the 1\%
precision quoted, e.g., in Refs.~\cite{20,21,23} corresponds to a 
68\% confidence level. To
obtain  good agreement between theory and experiment, it has been 
necessary to take into
account corrections to the Casimir force due to finite conductivity 
of the boundary metals,
surface roughness, and nonzero temperature
\cite{20,28,29,30,31,32,33,34}.

Although these new precision measurements of the Casimir force
were not especially designed or optimized to obtain stronger
constraints  on the predictions of
extra-dimensional physics, some of them were used for this
purpose \cite{34,35,36,37,38}.  This resulted in a strengthening of
previously known constraints by a factor as large as 4500
in some regions within the range $10^{-9}\,$m to $10^{-4}\,$m. This
means that Casimir force measurements have  become a
powerful competitor to both accelerator and gravitational experiments
in constraining  theoretical models of high energy physics. In
contrast to astrophysical and cosmological constraints, the
results of Casimir force measurements are less model dependent,
reproducible, and therefore more reliable.

In this paper we present detailed results of new, increased precision,
Casimir force measurements between a Cu-coated plate and an
Au-coated sphere (see Ref.~\cite{39} for preliminary data).
The use of a microelectromechanical torsional
oscillator (MTO) and of interferometric measurements of the
sphere-plate separations permitted  much higher
sensitivity to be achieved than in  previous Casimir force
experiments.  A careful error analysis has been performed, and
the experimental precision was determined at the
95\% confidence level. The
complete theory of the Casimir force, taking into account finite
conductivity and surface roughness corrections has been applied to the
experimental configuration. The finite conductivity corrections were
computed by the use of the Lifshitz formula \cite{51} and tabulated
optical data for the complex index of refraction. The surface roughness was
modelled using atomic force microscope (AFM) images of
the interacting surfaces. A comparison of the complete theory with
experimental data shows that they are in agreement over the whole
measurement range.  A minor disagreement at the shortest separations
noted in Ref.~\cite{39} is explained by the incomplete theory of
roughness corrections used in the earlier analysis.

Our results were sufficiently precise to shed light on the
temperature-dependence of the Casimir force.
Several theories predicting large  thermal corrections to the
Casimir force at small separations
\cite{40,41,42} were evaluated for our experiment. It is well known
that these corrections
disagree with the results obtained for ideal metals
in the framework of quantum field theory at nonzero
temperature in the Matsubara formulation.
Our experimental data
support the results of Refs.~\cite{31,32}, which are
consistent with the conclusions of thermal quantum field theory,
while ruling out the  existence of large thermal corrections at
small separations as proposed in Refs.~\cite{40,41} and in
Ref.~\cite{42}.

Finally, the experimental results presented here are used to
constrain the predictions of extra-dimensional physics in the
nanometer separation range.  The contributions from a Yukawa-type
hypothetical force have been calculated for our experimental configuration,
taking into account the effects of surface roughness.  The agreement between
theoretical and measured values of the Casimir force leads to the
strengthening of the known constraints on Yukawa forces by a factor
of up to 11 within the 56--330~nm interaction range.
In contrast to some previous constraints derived from
Casimir force measurements,  it is possible here to quantify the
confidence level of the obtained results.

This paper is organized as follows. In Sec.~II the experimental
configurations used for both static and dynamic measurements of
the  Casimir force are described.  Sec.~III presents the
experimental results with a discussion of their precision. In
Sec.~IV we calculate the Casimir force taking into account  all
relevant corrections. Sec.~V is devoted to the determination of the
theoretical precision, the comparison of experimental results with
theory, and to the evaluation of alternative methods for taking into
account thermal corrections. In Sec.~VI we use our results to obtain
constraints on  hypothetical forces predicted  by models of
extra-dimensional physics and extensions of the Standard Model.   We
conclude with Sec.~VII which summarizes all of our results.

\section{Experimental arrangement for static and dynamic measurements}

In our experiment, the Casimir force between two dissimilar metals
(gold and copper) was measured using a
MTO operating in both static
and dynamic modes. In the static regime the Casimir force
between an Au-coated sphere and a Cu-coated plate of the MTO was
measured.  In the dynamic regime the vertical separation between the
sphere and the plate was changed harmonically with time.  This leads
to a measurement of the $z$-derivative of the Casimir force, which is
equivalent to measuring the Casimir force per unit area, or the
Casimir pressure, for a configuration of two parallel
plates (see below).  Note that the dynamic measurement technique is 
used here to measure the
usual (static) Casimir effect. Hence, this measurement is unrelated to the
so-called dynamic Casimir effect which arises from the velocity
dependent Casimir forces or the creation of photons by the rapidly
oscillating plates \cite{34}.

When using mechanical oscillators to measure forces, one has to
confront the coupling of the oscillator with environmental
vibrations. Compared with cantilever oscillators, torsional
oscillators are less sensitive to mechanical vibrations that
induce a motion of the center of mass. Furthermore, the
miniaturization of the oscillators yields an improvement in its
quality factor and sensitivity \cite{25,43}. It is
consequently advantageous to use an electromechanical torsional
oscillator  to measure the Casimir force between two metals.

The experimental arrangement is shown schematically in Fig.~\ref{fig1}.
The MTO is made of a 3.5\,$\mu$m thick, 500\,$\times$ 500\,$\mu$m$^2$
heavily doped polysilicon plate suspended at two opposite points
by serpentine springs, as shown in the inset of
Fig.~\ref{circuit}. The springs are anchored to a silicon nitride
(SiN$_{x}$) covered Si platform. When no torques are applied the plate
is separated from the platform by a gap $\sim 2\,\mu$m. Two
independently contacted polysilicon electrodes located under the
plate are used to measure the capacitance between the electrodes and
the plate. For the MTO employed in this experiment, we calculated the
torsion coefficient $\kappa = (wt^3E_{si}/6L_{serp})
\simeq 9.5 \times 10^{-10}$Nm/rad \cite{44}, where
$w =2\,\mu$m
is the width of the serpentine, $t = 2~\mu$m is its thickness,
$L_{serp} = 500~\mu$m its length, and $E_{Si} = 180$ GPa is
Young's modulus for Si. This value is in good agreement with the
measured value $\kappa = 8.6 \times 10^{-10}$~Nm/rad.
The edges of the plate
are coated with 1\,nm of Cr followed by 200\,nm of Cu.
This layer of Cu constitutes one of the metals used in the
measurement of the Casimir force.

The remainder of the assembly, as shown in Fig.~\ref{fig1}, consists
of an
Au-coated Al$_{2}$O$_{3}$ sphere that can be brought in close
proximity to the Cu-coated plate. Al$_{2}$O$_{3}$ spheres with nominal
diameters ranging from 100\,$\mu$m to 600\,$\mu$m were coated with a
$\sim 1\,$nm layer of Cr followed by a $\approx 203\,$nm layer of
gold. The coated sphere used in the experiment was subsequently glued
with conductive  epoxy to the side of
an Au-coated optical fiber, establishing an electrical connection
between them. The sphericity of the Al$_{2}$O$_{3}$ balls, as measured
on a scanning electron microscope (SEM), was found to be within
the specifications of the manufacturer. As an example, a
600\,$\mu$m-diameter ball was found to have an ellipsoidal shape
with major and minor semi-axes of (298 $\pm$ 2)\,$\mu$m and (294.0
$\pm$ 0.5)\,$\mu$m, respectively. For the sake of clarity we will
refer to the ball as a sphere in the remainder of the paper. Deposition
induced asymmetries were found to be smaller than 10\,nm, the
resolution of the SEM. The entire setup (MTO and fiber-sphere) was
rigidly mounted into a can, where a pressure
$\lesssim 10^{-4}$\,torr was maintained. The can has built-in magnetic
damping vibration isolation and was, in turn, mounted onto an air table. This
combination of vibration isolation systems yielded base
vibrations with $\Delta z^{rms} < 0.05$\,nm for frequencies above
100\,Hz.

The fiber-sphere assembly was moved vertically by the combination of a
micrometer-driven and  a piezo-driven stage.
The MTO was mounted on a piezoelectric driven {\em xyz} stage which, in turn,
is mounted on a micrometer controlled {\em xy} stage. This combination
allows positioning the Au-coated sphere over the
Cu-coated plate. The separation
$z_i$ between the sphere and the Si-platform was controlled by the
$z$-axis of the {\em xyz} stage. A two color fiber interferometer-based
closed-loop system was used to keep $z_i$ constant. The error in the
interferometric measurements was found to be $\Delta z_i^{rms} =
0.32$\,nm, dominated by the overall stability of the closed-loop
feedback system. Since this error is much greater than the actual
mechanical vibrations of the system, the closed loop was turned off
while data acquisition was in progress.

The separation $z_{metal}$ between the two metallic surfaces (see
Fig.~\ref{fig1}) is given by $ z_{metal} = z_i - z_0 - z_g -
b\theta$, where $b$ is the lever arm between the sphere and the
axis of the MTO, and $\theta$ is the angle between the platform and
the plate ($\theta \ll 1$ has been used).  $z_0$ is the distance the
bottom of the sphere protrudes from the end of the cleaved fiber,
and $z_g$ includes the gap between the platform and the plate, the
thickness of the plate, and the thickness of the Cu layer. An
initial characterization of $z_0$, by alternately gently
touching the platform with the sphere and the bare fiber yielded
$z_0$ = (55.07 $\pm 0.07)~\mu$m. Also, $z_g$ = (5.73 $\pm$
0.08)\,$\mu$m,  was determined using an
AFM. Since errors in $z_g$ and $z_0$ propagate to $z_{metal}$,
it is necessary to provide a better characterization, which is
described below.

A  force $F(z)$ acting between the sphere and the plate produces
a torque  $\tau = bF(z) = \kappa\theta$ on the plate.
(In all cases
reported in this paper $\theta \leq 10^{-5}$~rad, so $\theta \ll 1$.)
Under these circumstances $\theta \propto \Delta C = C_{right} -
C_{left}$, where $C_{right}$ ($C_{left}$ ) is the capacitance between
the right (left) electrode and the plate  (Fig.~\ref{fig1}).
Consequently the force between the two metallic surfaces separated by a
distance $z$ is $F(z) = k\Delta C$, where $k$ is the
proportionality constant. The capacitance was measured using the circuit
schematically
shown in Fig.~\ref{circuit} \cite{45}. The 10~nV/Hz$^{1/2}$ electronic
noise of the amplifier stage is equivalent to an angular deviation
$\delta \theta \sim 10^{-9}$\,rad/Hz$^{1/2}$. This noise level is much
smaller than the thermodynamic noise at the measuring frequency $f$ = 10\,kHz
$\gg f_0 = 687.23$\,Hz \cite{46},

\begin{equation}
S_{\theta}^{1/2}=\left[{\frac{2k_B T Q}{\pi f_0\kappa}\left
(\frac{f_0}{f}\right )^2 }\right]^{1/2} \simeq 3 \times 10^{-7} {\rm
rad/Hz}^{1/2}, \label{noise}
\end{equation}

\noindent where $k_B$ is Boltzmann's constant, $T$ = 300\,K, and the
quality factor
$Q\simeq 8000$. (See the inset of Fig.~\ref{fig1}.) Using these values
a force sensitivity
$\delta F = \kappa S_{\theta}^{1/2}/b \simeq 1.4$\,pN/Hz$^{1/2}$
is obtained.

To provide a calibration of the proportionality constant $k$
between $F$ and $\Delta C$, we applied a known potential
difference between the Au-coated sphere and the Cu film.  This was
done at $z_{metal} > 3\,\mu$m, in configurations where the
contribution from the Casimir force is smaller than 0.1~\% of the
total force. Thus, the net force can be approximated by retaining only the
electrostatic force $F_e$ between a sphere and an infinite plane
\cite{47},

\begin{equation}
F_e = 2\pi\epsilon_0(V_{Au} - V_0)^2 \sum_{n=1}^{\infty}
\frac{[\coth (u) - n \coth (nu)]}{\sinh (nu)} \label{eq1}.
\end{equation}

\noindent Here $\epsilon_0$ is the permittivity of free space,
$V_{Au}$ is the voltage applied to the sphere, $V_0$ is the residual
potential difference between the metallic layers when they are
both grounded, and $\cosh u = [1+z/R]$, with $z =
z_{metal}+2\delta_0$ ($R$ is the radius of the sphere).
$2\delta_0$ is the average separation between the metal layers
when the test bodies come in contact (so that $z_{metal}=0$), and
is primarily determined by the roughness of the films.
Although the profile of the
roughness does not affect the value of the electric force
at large separations, its presence should be taken into account
in the determination of the separation between the
smoothed out surfaces.
In Eq.~(\ref{eq1}) it was found that only
the first two terms of the $z/R$ expansion gave a significant
contribution. Fig.~\ref{deltaV} shows the dependence of $\theta$
(and hence the electrostatic force) on the applied voltage $V_{Au}$. The
minimum in the force was found to occur when $V_0 = (632.5 \pm
0.3)$\,mV, which reflects the difference in the work functions of
the Au and Cu layers. This value was observed to be constant for
$z$ in the 0.2-5\,$\mu$m range and it did not vary when measured
over different locations in the Cu layer.

Once the value $V_0$ was found, Eq.~(\ref{eq1}) was used
to determine
three different parameters: (i) the proportionality constant
$k$ between the sphere-plate force and the measured difference in
capacitances; (ii) the radius $R$ of the coated sphere; and (iii) the
increase in the separation $2\delta_0$ between the two metallic
layers. Two of the curves obtained in the 3-5\,$\mu$m range are
shown in Fig.~\ref{fig2}. A simultaneous fit to more than 100 such curves
yields $k = (50280 \pm 6)$\,N/F, $R = (294.3 \pm 0.1)\,\mu$m, and
$\delta_0 = (39.4 \pm 0.3)$\,nm.

The above force sensitivity can be improved by performing a dynamic
measurement which directly uses the high quality factor of the MTO
\cite{25}.  In this approach, the separation between the sphere and the
oscillator was varied as $\Delta z_{metal} =A \cos(\omega_r t)$, where
$\omega_r$ is the resonant angular frequency of the MTO, and $A$ was
adjusted between 3 and 35\,nm for values of $z_{metal}$ between 0.2 and
1.2\,$\mu$m, respectively. The solution for the oscillatory motion
yields \cite{25},
\begin{equation}
\omega_r^2 = \omega_0^2 \left [
1-\frac{b^2}{I\omega_0^2}\frac{\partial F_C}{\partial z}\right ],
\label{approx}
\end{equation}
where $\omega_0 \simeq \sqrt{\kappa/I}$ for $Q \gg 1$,
$I\simeq 4.6 \times 10^{-17}$\,kg$\cdot$m$^2$
is the moment of inertia of the oscillator, and $F_C$ is the Casimir
force between the sphere and the plate.
Since $A \ll z_{metal}$, terms of higher order in
$\partial F_C/\partial z$
introduce a $\sim$ 0.1\% error at separations $z\geq 250\,$nm.
As before, Eq.~(\ref{eq1}) was
used to calibrate all constants.  We found
$\omega_0 =2\pi\times 687.23\,$Hz, and
$b^2/I = 1.2978 \times 10^9$\,kg$^{-1}$.
With an integration time of 10~s using a phase-lock-loop circuit
\cite{48}, changes in the resonant frequency of 10 mHz were
detectable.

The main source of error for the frequency measurement,
$\delta f=10\,$mHz,
is the error  in the trigger of the frequency meter. This
error $\delta t$  originated in the jitter of the signal due to
the relatively large thermodynamic noise-induced $\delta \theta$
observed at resonance \cite{46}. This dominant noise is

\begin{equation}
\delta f = \frac{\sqrt{16} f^{3/2} \delta t}{\sqrt{\Upsilon}},
\label{deltaf}
\end{equation}
\noindent
where $\Upsilon$ is the integration time.

Unlike the static regime where forces are measured, in the dynamic regime the
force gradient $\partial F_C/\partial z$ is  measured using
Eq.~(\ref{approx}) by observing the change in the resonant frequency
as the
sphere-plate separation changes.  According to the proximity force theorem
\cite{49,50},
\begin{equation}
F_C(z)=2\pi R E_C(z),
\label{eq5}
\end{equation}
\nn where $E_C(z)$ is the Casimir energy per unit area for two
infinitely large  parallel plates composed of the same materials as the
sphere and plate (see Sec.~IV for details). Differentiating Eq.~(\ref{eq5})
with respect to
$z$ one obtains
\begin{equation}
-\frac{\partial F_{C}(z)}{\partial z} = 2\pi R P_{C}(z),
\label{eq6}
\end{equation}
\nn where
$P_{C}(z)$ is the force per unit area between two infinite plates.
Thus, in the dynamic regime the Casimir force
gradient between the sphere and plate can be directly related to the Casimir
pressure between two infinite parallel plates composed of the same
materials.  Since  the formula for the Casimir pressure for the parallel
plate configuration is readily obtained, it proves more convenient to use
Eq.~(\ref{eq6}) to express the results for the dynamic measurements in terms
of the parallel plate pressure
$P_{C}$ instead of the force derivative $\partial F_{C}/\partial z$.
{}From
Eq.~(\ref{deltaf}), the error in the measurement of the frequency translates
into  an equivalent pressure sensitivity given by
$\delta P_{C} \simeq 4 \times 10^{-4}$~Pa/Hz$^{1/2}$.

Finally, one additional test was performed, in this case to analyze the
influence of the finite extent of the Cu layer on the measured
forces. The analysis was done using, once again, the electrostatic
force given by Eq.~(\ref{eq1}). Fig.~\ref{bandz} shows the
relevant data. When the sphere was moved parallel to the axis of
the oscillator over 20\,$\mu$m no change in $\theta$ was observed.
When the motion was instead perpendicular to the axis of the
oscillator the dependence expected from Eq.~(\ref{eq1}) was
obtained within the experimental error. We thus conclude that
there are no significant deviations from the assumption that the
Cu plane is of infinite extent.

\section{Experimental results and their precision}

\subsection{Surface roughness of samples}

The electric force measurements used to calibrate the apparatus
were performed at large separations so that the Casimir force
could be neglected, and the plate-sphere
separation was calibrated between the middle levels of the surface
roughness.
By contrast, the Casimir force at small separations
depends sensitively on the profile of the surface roughness, and hence
the surface roughness should be
carefully analyzed and characterized. The
topography of the metallic films was investigated using an
AFM  probe with a radius of curvature
$r_c=5\,$nm in tapping mode.  Regions of the metal
plate and the sphere varying in size from
$1\,\mu\mbox{m}\times 1\,\mu$m to $10\,\mu\mbox{m}\times 10\,\mu$m
were scanned. A typical surface scan of a
$1\,\mu\mbox{m}\times 1\,\mu$m region is shown in Fig.~\ref{AFM},
where the lighter tone corresponds to higher regions. As seen in
Fig.~\ref{AFM}, the major distortions are the large
mounds situated irregularly on the surface.  Also noticeable in
Fig.~\ref{AFM} are streaks which arise from high frequency noise
with amplitude $h^{rms}\sim 1\,$nm. This noise is caused by the
oscillation of the free-standing MTO while acquiring AFM images.

In order to include the effects of surface roughness in the Casimir
force calculations, the fraction of the surface area $v_i$ with height
$h_i$ is needed.  Data resulting from the
most representative scan of  a $10\,\mu\mbox{m}\times 10\,\mu$m
region are shown in Fig.~\ref{profile}. The heights $h_i$ are plotted
along the vertical axis as a function of the fraction $w_{i}$  of the
total surface area having height less than $h_{i+1}$. The width of
each horizontal step is equal to the fraction of the total area $v_i$ with
heights $h_i\leq h <h_{i+1}$. For example,
regions with heights $h<h_2=8.2\,$nm are referred to the
first (bottom) distortion level $h_1=0$ with $v_1=0.0095$,
and regions with heights $h_2\leq h<h_3=9.87\,$nm
are referred to the second distortion level $h_2$ with $v_2=0.0121$.
Regions with heights $h\geq h_{59}=98.5\,$nm are
referred to $h_{59}$, and occupy a fraction $v_{59}=0.085$
of the total area. Evidently $w_i=v_1+v_2+\ldots+v_i$, and
$w_{59}=1$.

The data of Fig.~\ref{profile} will be used in Sec.~IV for the
computation of the Casimir force taking into account roughness corrections
(in Ref.~\cite{39} a simplified model of roughness was
used).
These data are also required for the precise determination of the so-called
zero roughness level $H_0$ relative to which the mean value of the function,
describing the total roughness, is zero. For convenience in comparison with
theory, all separation distances in the Casimir force measurements presented
below are measured between the zero roughness levels.

The zero roughness level $H_0$ is found using the equality
\begin{equation}
\sum\limits_{i=1}^{59}(h_i-H_0)v_i=0.
\label{eq7}
\end{equation}
\nn By combining the data of Fig.~\ref{profile} with Eq.~(\ref{eq7}), one
finds
$H_0\approx 35.46\,$nm. It is seen that the zero roughness level
is slightly different from the quantity $\delta_0\approx 39.4\,$nm,
the correction to the separation on contact determined by the results
of the electric force measurements (see Sec.~II). The separation
distances between zero roughness levels $z=d-2H_0$ used below
($d$ is the separation between the bottom roughness levels) are
larger by $2(\delta_0-H_0)=7.88\,$nm than the separations
$d-2\delta_0$ defined by the electric force measurements.
(The difference
between $H_0$ and $\delta_0$ can be explained by a minor modification
of the highest roughness peaks before the beginning of the
Casimir force measurements.)

\subsection{Static measurements of the Casimir force}

To measure the Casimir force in the static regime, the bridge
(schematically shown in Fig.~\ref{circuit}) was first balanced using
two identical capacitors replacing the MTO.
Then the MTO was put back in place and the voltages $V_1$ and
$V_2$ adjusted to give a null signal. This last adjustment is required
to take into account residual asymmetries in the MTO. It
was found that the difference between $V_1$ and $V_2$ corresponded
to a variation in $z$, $\delta z \leq 6\,$nm. Once the MTO was
mounted and the can was evacuated, the sphere was brought into
proximity with the MTO and the electrostatic measurements were performed.
Without breaking the vacuum in the system the Casimir force
measurements were then carried out. Fig.~\ref{force} shows one
such data set out of 19 runs.
   Each data point was  obtained with an integration
time of 10\,s, a time interval which represents a good compromise for
the $\approx 300$ data points taken per run. It is worth mentioning,
however, that the force sensitivity can be improved by using
longer integration times.

The following analysis of the experimental precision is based on all $n=19$
series of measurements. For this purpose we calculate the mean values of the
measured Casimir force

\begin{equation}
\bar{F}_C^{exp}(z_m)=\frac{1}{n}
\sum\limits_{i=1}^{n}{F}_{C,i}^{exp}(z_m)
\label{eq8}
\end{equation}

\nn at different separations $z_m$ within the measurement separation
range from 190\,nm to $\approx~1.15\,\mu$m. The mean square error of
$\bar{F}_C^{exp}$ is equal to

\begin{equation}
s_n(z_m)=\left\{\frac{1}{n(n-1)}
\sum\limits_{i=1}^{n}\left[\bar{F}_C^{exp}(z_m)-
{F}_{C,i}^{exp}(z_m)\right]^2\right\}^{1/2}.
\label{eq9}
\end{equation}

\nn Our calculations show that $s_n(z_m)$ does not depend sensitively on $m$.
The largest value
$s_n=0.143\,$pN  can be taken as the value of the
mean square error of $\bar{F}_C^{exp}$
within the whole measurement range. Taking into
account that at $\alpha=95$\% confidence level the
Student's coefficient is $t_{\alpha,n}=2.1$,
one obtains for the random absolute error of the Casimir force
measurements in the static regime

\begin{equation}
\Delta_{st}^{\!\!{rand}}F_C^{exp}=s_n t_{\alpha,n}
\approx 0.3\,\mbox{pN}.
\label{eq10}
\end{equation}

\nn In fact this effectively gives the total absolute error
$\Delta_{st}^{\!\!{tot}}F_C^{exp}$ since the systematic errors
are far below 0.1\,pN for an integration time of 10\,s.
As a result, at the shortest separation of about 188\,nm the
relative error of the Casimir force measurement is  0.27\%.
We note that the true value of the Casimir force at a separation
$z$ lies in  the confidence interval

\[
\left[{\bar{F}}_C^{exp}(z)-\Delta_{st}^{\!\!{tot}}F_C^{exp},
{\bar{F}}_C^{exp}(z)+\Delta_{st}^{\!\!{tot}}F_C^{exp}\right]
\nonumber
\]

\nn with a probability of 95\%.

\subsection{Dynamic measurements of the Casimir pressure}

The results for the parallel plate Casimir pressure in the
dynamic regime are shown in Fig.~\ref{pressure}, which presents one data
set of 5 runs. Although the measurement was extended down to
a separation of $\sim$ 180\,nm, only data for separations above
260\,nm are plotted. Data points below 260\,nm show effects of
nonlinear behavior of the oscillator \cite{25}.

The error analysis is performed in the same way as for the static
regime. For $l=5$ runs, with  $\approx 300$ points per run,
the largest mean square error is found to be $s_l=0.11\,$mPa.
The Student's coefficient at $\alpha=95$\% confidence level
is $t_{\alpha,l}=2.8$ leading to a random absolute error
$\Delta_{dyn}^{\!\!{rand}}P_C^{exp}=0.31\,$mPa.
In the case of dynamic measurements, however, it is not possible
to neglect systematic errors compared to random errors.
As noted in Sec.~II, there is an error
$\delta\omega\approx 2\pi\times 10^{-2}\,$Hz which
from Eq.~(\ref{approx}) leads to the absolute error in
$\partial F_C/\partial z$  $\approx 4.2\times 10^{-7}\,$N/m.
The latter, when combined with the error associated with the
sphere's radius (equal to
$0.1\,\mu$m), leads via Eq.~(\ref{eq6}) to
a systematic error on the pressure
$\Delta_{dyn}^{\!\!{syst}}P_C^{exp}$, which varies from 0.31\,mPa at
the shortest separation $z=260\,$nm to 0.23\,mPa at all
separations $z\geq 450\,$nm.
As a result, the total absolute error of pressure measurements
in the dynamic regime is equal to

\begin{equation}
\Delta_{dyn}^{\!\!{tot}}P_C^{exp}=
\Delta_{dyn}^{\!\!{rand}}P_C^{exp}+
\Delta_{dyn}^{\!\!{syst}}P_C^{exp}.
\label{eq11}
\end{equation}

\nn This error is $z$-dependent and varies from 0.62\,mPa at
$z=260\,$nm to 0.54\,mPa at $z\geq 450\,$nm.
Hence,  the relative error of the Casimir
pressure measurement at the shortest separation of about
260\,nm is 0.26\%.

\section{Theoretical determination of the Casimir force}

\subsection{Casimir force and pressure including finite conductivity}

As described in Section II, the static and dynamic measurements were
carried out using a Au-coated sphere over a Cu-coated plate. The thicknesses
of both metal coatings were much larger than the plasma
wavelength $\lambda_p$ of both metals so that we can calculate the Casimir
force as if the sphere and plate were composed of solid Au and Cu
respectively.  As a first approximation, we consider the plate to have an
infinite area.  (Corrections due to the finite size of the plate
will be estimated below.)  The Casimir force between an infinite plate and a
sphere (which was measured in the static regime, see Sec.~II) can then be
found using the Lifshitz formula for two plane parallel plates and the
proximity force theorem \cite{51,52}

\begin{equation}
F_C(z)=\frac{R}{2\pi}\int_{0}^{\infty}k_{\bot}dk_{\bot}
\int_0^{\infty}d\xi\left\{\ln\left[1-r_{\|}^{(1)}(\xi,k_{\bot})
r_{\|}^{(2)}(\xi,k_{\bot})e^{-2qz}\right]\right. \label{eq12} \end{equation}
\[
\left.\phantom{aaaaaaaaa}
+\ln\left[1-r_{\bot}^{(1)}(\xi,k_{\bot})
r_{\bot}^{(2)}(\xi,k_{\bot})e^{-2qz}\right]\right\}.
\nonumber
\]

\nn Here $q^2=k_{\bot}^2+\xi^2$, $k_{\bot}$ is the modulus of the
wave vector in the plane of the plates, and $r_{\|,\bot}^{(l)}$
($l=1,\,2$ for Cu, Au, respectively) are the reflection
coefficients for two independent polarization states computed
along the imaginary frequency axis $\omega=i\xi$. We note
that the errors introduced by the proximity force theorem used to
derive Eq.~(\ref{eq12}) are smaller than $z/R$ \cite{53,54}.  This is
a correction of less than 0.06\% at the shortest separation distance, where
the force measurements are most precise. (Recall that the
experimental precision at $z = 188$\,nm was found to be 0.27\%.)
Eq.~(\ref{eq12}) takes into account the finite conductivity
corrections to the Casimir force due to real metal boundaries, but
does not consider the effect of surface roughness and thermal
corrections which will be treated later.

The reflection coefficients used in Eq.~(\ref{eq12}) can be
represented in terms of either the dielectric permittivity or the
surface impedance. In terms of the dielectric permittivity, as in
the original Lifshitz formula (denoted by the subscript
$L$), the reflection coefficients are given by

\begin{equation}
r_{\|,L}^{(l)}(\xi,k_{\bot})=
\frac{k^{(l)}-\varepsilon^{(l)}(i\xi)q}{k^{(l)}+\varepsilon^{(l)}(i\xi)q},
\quad r_{\bot,L}^{(l)}(\xi,k_{\bot})= \frac{q-k^{(l)}}{q+k^{(l)}},
\label{eq13}
\end{equation}

\nn where
${k^{(l)}}^2=k_{\bot}^2+\varepsilon^{(l)}(i\xi)\xi^2$. In terms of
the surface impedance, the reflection coefficients are given by \cite{33}

\begin{equation}
r_{\|}^{(l)}(\xi,k_{\bot})=
\frac{Z^{(l)}(i\xi)\xi-q}{Z^{(l)}(i\xi)\xi+q}, \quad
r_{\bot}^{(l)}(\xi,k_{\bot})=
\frac{Z^{(l)}(i\xi)q-\xi}{Z^{(l)}(i\xi)q+\xi},
\label{eq14}
\end{equation}

\nn where $Z^{(l)}(i\xi)$ is the impedance for Cu or Au
computed along the imaginary frequency axis.

It has been shown recently \cite{33} that for real metals at nonzero
temperature and interacting at relatively large separations,
the representation described in Eq.~(\ref{eq14}) is
preferable. It is free of contradictions with thermodynamics, which
arise when reflection coefficients expressed by Eq.~(\ref{eq13})
are used in combination with the Drude dielectric function
containing a nonzero relaxation parameter \cite{55}.  Within the
experimental separation ranges of Sec.~II (0.2--1.2 $\mu$m) the characteristic
angular frequency of the Casimir effect $\omega_c=1/(2z)$ lies within the
region of infrared optics (using $c=1$),
where the dielectric permittivity of the plasma
model and the surface impedance along the imaginary frequency axis are given
by

\begin{equation}
\varepsilon(i\xi)=1+\frac{\omega_p^2}{\xi^2}, \quad
Z(i\xi)=\frac{\xi}{\sqrt{\omega_p^2+\xi^2}}.
\label{eq15}
\end{equation}

\nn As  demonstrated in \cite{56}, both formulations of the reflection
coefficients,  Eqs.~(\ref{eq13}) and  (\ref{eq14}), lead to the same
computational results for the
Casimir force, Eq.~(\ref{eq12}), when  Eq.~(\ref{eq15}) is taken into
account.

In the dynamic regime of Sec.~II, the Casimir force gradient between the
sphere and plate was measured.  As shown above, this force gradient can be
re-expressed in terms of the Casimir pressure acting between two
parallel plates composed of the same materials.   This Casimir pressure
between parallel plates can also be represented in terms of the
reflection coefficients

\begin{equation}
P_C(z)=-\frac{1}{2\pi^2}\int_{0}^{\infty}k_{\bot}dk_{\bot}
\int_0^{\infty}qd\xi\left\{\left[\frac{e^{2qz}}{r_{\|}^{(1)}(\xi,k_{\bot})
r_{\|}^{(2)}(\xi,k_{\bot})}-1\right]^{-1}\right. \label{eq16} \end{equation}
\[
\left.\phantom{aaaaaaaaaaaaaaaa}
+\left[\frac{e^{2qz}}{r_{\bot}^{(1)}(\xi,k_{\bot})
r_{\bot}^{(2)}(\xi,k_{\bot})}-1\right]^{-1}\right\}.
\nonumber
\]

\nn For separations $z>\lambda_p$ the Casimir pressure can
be computed using either set of reflection coefficients,
Eq.~(\ref{eq13}) or Eq.~(\ref{eq14}), with $\varepsilon(i\xi)$ or $Z(i\xi)$
given by Eq.~(\ref{eq15}). At the shortest separations
$z<\lambda_p$ the impedance at characteristic frequencies is not
small,  and Eq.~(\ref{eq13}) should be
used to calculate both the Casimir pressure and force. The most
accurate results at these separations are obtained by using
tabulated data for the imaginary part of the dielectric
permittivity \cite{57}. These data are substituted into the
dispersion relation

\begin{equation} \varepsilon(i\xi)=1+\frac{2}{\pi}
\int_0^{\infty} \frac{\omega
\mbox{Im}\varepsilon(\omega)}{\omega^2+\xi^2}d\omega
\label{eq17}
\end{equation}

\nn to obtain the dielectric permittivity along the imaginary
frequency axis. The Casimir force, Eq.~(\ref{eq12}),
and pressure,
Eq.~(\ref{eq16}), can then be calculated as in Refs.~\cite{28,29}. This
procedure was used here to calculate $F_C(z)$ and $P_C(z)$. The
available tabulated data \cite{57} were extended using
the Drude model with the following plasma frequencies and
relaxation parameters \cite{28}: $\omega_p^{(1)}=9.05\,$eV,
$\omega_p^{(2)}=9.0\,$eV, $\gamma^{(1)}=30\,$meV,
$\gamma^{(2)}=35\,$meV, for Cu and Au, respectively. At large
separations $z>\lambda_p$ our results almost coincide with those
obtained in the framework of the plasma model given by
Eq.~(\ref{eq15}). Note that at smaller separations $z<\lambda_p$
our results are practically independent of the chosen
extrapolation and are completely determined by the available
tabulated data.

It is useful to compare the calculated results for the real metals used
with the ideal case when both sphere and plate were composed of a perfect
metal with infinite conductivity.   For example, the ratios of the calculated
Casimir force to the force between an ideal sphere and plate are
0.467, 0.544, and 0.842 at separations $z=70\,$nm, 100\,nm, and 500\,nm,
respectively.   Similarly,  the ratios of the calculated
Casimir pressure between real plates to the pressure between ideal plates
are 0.393, 0.468, and 0.799 for the same separations. These results are very
close to those computed for Au-Au and Cu-Cu
\cite{28,29}, which can be  explained by the similarities of the
optical data for Cu and Au.

\subsection{Casimir pressure and force including surface roughness}

As mentioned previously, the theoretical results obtained from
Eqs.~(\ref{eq12}) and (\ref{eq16}) take into account the finite
conductivity of the boundary metals but neglect the effect of
surface roughness. This effect, however, may constitute a correction of
several tens of percent at the shortest separations depending on the
character of the roughness. Hence, a precise computation of the Casimir
force requires a careful characterization of the roughness as was performed
in Sec.~IIIA.

The roughness correction to the Casimir interaction is computed using the AFM
images of the surfaces, like the one shown in Fig.~\ref{AFM}. As seen from
Fig.~\ref{AFM}, the characteristic longitudinal scale of the surface
roughness is larger than the surface separation (especially in the region of
the shortest separations where the effect of roughness is most significant).
In this case the additive method \cite{34,58} can be used to calculate the
Casimir force taking  account of roughness. As a result,
the Casimir pressure between two plates with roughness
corrections taken into account is given by

\begin{equation}
P_C(z)=\sum\limits_{i,j=1}^{n}v_iv_jP_C(z+2H_0-h_i-h_j),
\label{eq18}
\end{equation}

\nn where $v_{i}$ is defined in Sec.~IIIA,  $P_C(z)$ is given by
Eq.~(\ref{eq16}), and the index
$i$ relates to one plate and $j$ to the other.  A plot of the relief
heights $h_i$ ($i=1,2,\ldots,n=59$) on the plate versus the
fraction of the plate area $w_i$ with height $h<h_{i+1}$ is shown
in Fig.~\ref{profile}. It should be remembered that
$H_0=35.46\,$nm is the zero roughness level from which all
separations $z$ are measured (see Sec.~IIIA).
Note also that the separation distances $z+2H_0-h_i-h_j$ in
Eq.~(\ref{eq18}) may be much smaller than $\lambda_p$ for large
relief heights $h_i,\,h_j$. Hence, one should use
the optical tabulated data when calculating the finite
conductivity corrections.

In a similar manner, the Casimir energy
for  two parallel plates with roughness
is given by Eq.~(\ref{eq18}) after substituting $E_C$
for $P_C$.  It then follows that by the use of the  proximity
force theorem, the Casimir force for a sphere
above a plate, accounting for  both roughness and finite
conductivity corrections,  is given by

\begin{equation}
F_C(z)=\sum\limits_{i,j=1}^{n}v_iv_jF_C(z+2H_0-h_i-h_j),
\label{eq19}
\end{equation}

\nn where $F_C(z)$ is given by Eq.~(\ref{eq12}).

We note that  Eqs.~(\ref{eq18}) and (\ref{eq19}) describe not only
the separate effects of finite conductivity and surface roughness
on the Casimir pressure and force,  but also their combined effect.
This is especially
important at the shortest separations where the
corrections are not small and cannot be represented
as a product of two separate factors, one each for roughness and finite
conductivity.

\section{Comparison of Theory with Experiment and tests of alternative
thermal corrections}

The theoretical Casimir force $F_C(z)$ acting between a sphere and
a plate was computed using Eq.~(\ref{eq19}) for all separations where
it was measured (19 sets of measurements containing $\approx 300$
experimental points each). In Fig.~\ref{deltaF} the difference between
the theoretical and experimental force values

\begin{equation} \Delta
F_C(z_i)=F_C^{th}(z_i)-F_C^{exp}(z_i)
\label{eq20}
\end{equation}

\nn as a
function of surface separation $z_i$ is presented for one set of
measurements. As can be seen from the figure, the values of $\Delta
F_C(z_i)$ are clustered around  $\Delta F_C(z_i) = 0$,
demonstrating good agreement between theory and experiment.

To quantify the level of agreement between theory and experiment
we consider the root mean square (r.m.s.) deviation  $\sigma_{N}^{F}$
defined as

\begin{equation} \sigma_N^F=\left\{\frac{1}{N}
\sum_{i=1}^{N}\left[F_C^{th}(z_i)-F_C^{exp}(z_i)\right]^2\right\}^{1/2},
\label{eq21}
\end{equation}

\nn where $N$ is a number of points under
consideration.

For example, if the first 250 points from all 19 sets of static
measurements are considered (separations larger than 1\,$\mu$m
are not considered due to the large experimental relative error),
one obtains $N=4750$, and $\sigma_{4750}^{F}\approx 0.6\,$pN,
which is less than the theoretical error (see below) but
two times larger than the absolute error of the force measurements
given by Eq.~(\ref{eq10}). The r.m.s. deviation depends slightly
on the separation region under consideration. Thus, if separations
$z\geq 400\,$nm are considered, the first 185 points of all sets
of measurements lead to $N=3515$, $\sigma_{3515}^{F}\approx
0.45\,$pN. For one set of measurements shown in Fig.~\ref{deltaF},
$N=250$ and $\sigma_{250}^{F}\approx 0.66\,$pN (close to the above
value when all sets of measurements are considered). It can
also be seen from Fig.~\ref{deltaF} that there is a slight shift of the
mean difference force value below the zero level equal to $-0.045$\,pN,
whose magnitude is much smaller than the absolute error of the force
measurement.

A similar analysis was performed for the parallel plate Casimir
pressure obtained from the dynamic measurement. The difference
between the theoretical and experimental pressures
$\Delta P_C(z_i)$ is defined as in Eq.~(\ref{eq20}), and the r.m.s.
deviation $\sigma_N^P$ is defined as in Eq.~(\ref{eq21}) with the
substitution of $F_C$ for $P_C$. Considering the first 235 points of all 5
sets of dynamic measurements one obtains $N=1175$, and
$\sigma_{1175}^{P}\approx 0.5\,$mPa. Once again, the r.m.s.
deviation depends slightly on the separation interval. Thus, for
$310\,\mbox{nm}\leq z\leq 420\,$nm (35 points), $N=175$ and
$\sigma_{175}^{P}\approx 0.44\,$mPa if all five sets of
measurements are considered. For $z>310\,$nm ($220\times 5=1100$
points), $\sigma_{1100}^{P}\approx 0.34\,$mPa.

In Fig.~\ref{deltaP}, the quantity $\Delta P_C(z_i)$ is presented for
one set of measurements ($N=235$). Here the shift of the mean
difference pressure value below zero is equal to $-0.26$\,mPa
(whose magnitude is less than the absolute error of pressure measurements).
The r.m.s. deviation between theory and experiment in Fig.~\ref{deltaP}
is $\sigma_{235}^{P}\approx 0.43\,$mPa, i.e., even less than for all
five sets of dynamic measurements. Both Figs.~\ref{deltaF} and \ref{deltaP}
demonstrate good agreement between theory and experiment.

To estimate the theoretical precision, we consider  other effects,
in addition to finite conductivity and surface roughness,
which were not taken into account in  Eqs.~(\ref{eq18}), (\ref{eq19}) for
the Casimir pressure and force.
Most prominent among these are finite-temperature
corrections, which  will be dealt with in detail later
in this section.
Another correction which should be  considered arises from the fact that
the plate used in the experiment is not infinite in extent.
For the plate used,
whose dimensions are $500\times 500\,\mu\mbox{m}^2$, this correction is
evidently negligible if the sphere's center is located above the mid-point of
the plate.  However, in the configuration shown in Fig.~\ref{fig1}, the
sphere's center is above a point displaced from the right plate
boundary by a distance $L=50\,\mu$m.   In this case, the Casimir force
should be multiplied by a correction factor $\beta$ that was found in
Ref.~\cite{59}

\begin{equation}
\beta(z)=1-\frac{z^3}{R^3}\left(1- \frac{1}{\sqrt{1 +
L^{2}/R^{2}}}\right)^{-3},
\label{eq22}
\end{equation}

\nn where $R$ is the radius of the sphere.  In our case, at the largest
separation $z=1\,\mu$m, this leads to $\beta=0.9865$.  However, taking
into account the fact that only one of the four sides of the plate is
close to the sphere center projection reduces the correction factor to
$\beta=0.997$. Thus, the correction is in fact less than 0.3\%, and at a
separation $z=500\,$nm it is less than 0.04\%. The same considerations apply
to the pressure between two parallel plates. Consequently, the
correction to the theoretical force and pressure due to the
finiteness of the plate is much smaller than the uncertainty
introduced by the sample-to-sample variation of the optical data.
These sample-dependent variations in the index of refraction may
lead to an error of approximately 1\% \cite{28,29}.
An even smaller uncertainty is introduced into the computation by the
effect of roughness when the topography of the
surface is carefully characterized.  A recently discussed correction 
due to the surface
plasmon \cite{59a} is also
not significant. The surface plasmon propagates when the frequency of the
electromagnetic wave is greater than $\omega_{p}$. Recall that in our
case the shortest separations are 190~nm and 260~nm in the static and dynamic
regimes, respectively. This leads to the highest characteristic
frequencies $\omega_{c} = 0.8 \times 10^{15}$ rad/s and  $0.6 \times
10^{15}$ rad/s, respectively, which are 17 and
23 times, respectively,  less than the plasma frequency for Au and
Cu. As a result, the
correction due to the surface plasmon in our case is much less than 1\%
even at shortest separations used.

As mentioned earlier, an important correction which may influence the
magnitude of the Casimir interaction is due to the effect of nonzero
temperature. This has been the subject of considerable controversy during the
last few years (see, e.g., \cite{31,32,33,34,40,41,42,55,60}). For an ideal
metal the thermal correction can be determined using the Matsubara
formulation of thermal quantum field theory (TQFT)\cite{34,61,62,63,64,65}.
The question is whether the thermal correction for good (but real)
conductors is small at small separations, as is the case for ideal metals,
or does it differ qualitatively from that for ideal metals.
The approach of Refs.~\cite{31,32}, which we
call ``traditional'' since it yields results consistent with earlier studies
of thermal effects, leads to a qualitatively identical thermal correction for
real and ideal metals, as given by TQFT. For example, in the configuration of
two parallel plates made of Cu and Au, the Casimir pressures at $z=300\,$nm
and $z=500\,$nm at zero temperature are equal to
$-136$\,mPa and $-17.0$\,mPa, respectively.  The traditional thermal
corrections \cite{31,32} at room temperature $T=300\,$K
for these cases are equal to $-0.00863$\,mPa and
$-0.00441$\,mPa, respectively. (For comparison, in the case of ideal metals,
the corresponding Casimir pressures at $T=0\,$K are equal to
$-160.43$\,mPa and $-20.79$\,mPa, respectively, and the thermal
correction at $T=300\,$K is $-0.00204$\,mPa and is independent of
separation.) It is clearly seen that the traditional thermal
corrections at room temperature
are very small, and even the improved sensitivity of
our experiment is not sufficient to measure them.  Their contribution
to the Casimir pressure is $\sim0.006$\% at $z=300\,$nm and
$\sim0.03$\% at $z=500\,$nm, and can therefore be neglected.

The situation with the alternative thermal corrections,
proposed in Refs.~\cite{40,41,42}, is quite different. These
corrections at separations $z>\lambda_p$
are much greater than those predicted by the traditional approach. According
to Refs.~\cite{40,41}, in the case of two parallel plates made of
real metals there is an additional thermal correction linear in
temperature given by

\begin{equation} \Delta_TP_C^{(1)}(z)=\frac{k_BT}{16\pi
z^3} \int_{0}^{\infty}y^2dy\left[
\frac{e^y}{r_{\bot}^{(1)}(0,y)r_{\bot}^{(2)}(0,y)}-1 \right]^{-1},
\label{eq23}
\end{equation}

\nn where

\begin{equation} r_{\bot}^{(l)}(0,y)=
\frac{y-\sqrt{y^2+4z^2{\omega_p^{(l)}}^2}}{y+
\sqrt{y^2+4z^2{\omega_p^{(l)}}^2}}.
\label{eq24}
\end{equation}

\nn The effect of this correction is not as small as for the traditional
approach.  For example, at separations $z=300\,$nm and 500\,nm
Eq.~(\ref{eq23}) leads to
$\Delta_TP_C^{(1)}=4.89\,$mPa and $\Delta_TP_C^{(1)}=1.23\,$mPa,
respectively, at $T=300\,$K,
i.e., to 3.6\% and 7.24\%, respectively, of the total Casimir
pressure.  Since these values far exceed the errors of the present
experiment, the new results can be used as a decisive experimental
test for the theoretical predictions made in Refs.~\cite{40,41}.

In Fig.~\ref{T1} the difference between the theoretical and experimental
Casimir pressures $P_C^{th,1}-P_C^{exp}$ is presented as a function of
separation for the same set of measurements as used in Fig.~\ref{deltaP}.
Here, however, the theoretical pressure $P_C^{th,1}$
is computed using
the alternative thermal correction given by Eq.~(\ref{eq23}). It is
obvious that at separations $\lesssim 700$\,nm
the quantity $P_C^{th,1}-P_C^{exp}$
deviates significantly from zero.
At the shortest separation $z=260\,$nm this deviation reaches
5.5\,mPa. Thus, the linear thermal correction to the Casimir
pressure proposed in Refs.~\cite{40,41} is ruled out by the
present experimental results.  (Note that in recent preprint 
\cite{65a} qualitative arguments
are presented on the role of the finite size of the plate and finite
thickness of the gold layer. According to \cite{65a} these effects could
lead to a 25\%
discrepancy between the experiment of Ref.~\cite{19}, and the proposed
alternative thermal correction \cite{40,41} at a separation 1 $\mu$m.)

A second alternative thermal correction to the Casimir
pressure between real metals was proposed in Ref.~\cite{42}.
For our case of two metal plates it is expressed as

\begin{equation}
\Delta_TP_C^{(2)}(z)=-\frac{k_BT}{8\pi z^3}\zeta(3)
+\Delta_TP_C^{(1)}(z),
\label{eq25}
\end{equation}

\nn where $\zeta(x)$ is the Riemann zeta-function,  and
$\Delta_TP_C^{(1)}(z)$ is defined in Eq.~(\ref{eq23}).
This correction at $T=300\,$K is also much larger
than the traditional one for the separations used in our experiment.
At the separation $z=300\,$nm Eq.~(\ref{eq25}) leads to
$\Delta_TP_C^{(2)}=-2.44\,$mPa, i.e., to a 1.8\% correction to the total
Casimir pressure (the experimental precision at this separation is
0.43\%). Thus, the present experiment also provides a test for the
theoretical predictions of Ref.~\cite{42}.

In Fig.~\ref{T2} the difference between the theoretical and
experimental Casimir pressures $P_C^{th,2}-P_C^{exp}$ is presented
as a function of separation for the same set of measurements as in
Fig.~\ref{deltaP}. Unlike Fig.~\ref{deltaP}, the theoretical
pressure $P_C^{th,2}$ is computed using the
second alternative thermal correction given by Eq.~(\ref{eq25}).
As seen from Fig.~\ref{T2},
   at separations less than 600\,nm
the quantity $P_C^{th,2}-P_C^{exp}$
   deviates significantly from zero, and reaches 5\,mPa
at a separation $z=260\,$nm.
It follows that the thermal correction of Ref.~\cite{42} is also in
contradiction with the results of the present  experiment.

The conclusion that the alternative approaches to thermal corrections to the
Casimir force do not agree with our experimental results is not
surprising upon recognizing that the approaches of Refs.~\cite{40,41,42}
violate the Nernst heat theorem \cite{55}. To correctly describe the
influence of thermal effects on the Casimir force between real (i.e.,
non-ideal) metals requires a proper understanding of the zero-frequency
contribution in the Lifshitz formulas given by Eqs.~(\ref{eq12}) and
(\ref{eq16}). When the Drude dielectric function with a nonzero relaxation
parameter is substituted into Eq.~(\ref{eq13}), the approach of
Refs.~\cite{40,41} follows and the thermal correction given by
Eq.~(\ref{eq23}) is found. If one then modifies the reflection coefficient
$r_{\bot}(0,k_{\bot})$ by setting it equal to unity (as for real photons),
the thermal correction (\ref{eq25}) is obtained \cite{42}. However, to avoid
contradictions with fundamental physical principles and experiment, it is
necessary to start from the physical behavior of the surface impedance
in the appropriate range of characteristic frequencies (i.e., infrared
optics in our case), and to extrapolate to zero frequency \cite{33}.
Under these conditions, the traditional thermal correction
between real metals is recovered, and this thermal correction is in
agreement with both thermodynamics and the present experiments. Furthermore,
it  transforms smoothly into the limiting case of ideal metals
as described by the Matsubara
formulation of TQFT.

To conclude this section, the theoretical uncertainty, which is
approximately 1\% of the Casimir force or pressure, exceeds the
experimental uncertainty at the shortest separations used in this
experiment (0.27\% for the force at 188\,nm and 0.26\% for the pressure at
   260\,nm). However, at
separations $z> 370\,$nm the experimental uncertainties exceed the
theoretical uncertainties.  Within the achieved levels of precision there is
good agreement between theory and experiment over the entire
measurement range.

\section{Constraints on New Yukawa Forces and extra-dimensional physics}

As mentioned in the Introduction, in many extensions to the Standard model,
including theories with  large compact extra dimensions \cite{1},  the
potential energy between two point masses $m_1$ and $m_2$  separated by a
distance $r$  is given by the usual Newtonian potential with a Yukawa
correction \cite{1,12,13},

\begin{equation} V(r)=-\frac{Gm_1m_2}{r}\left( 1+\alpha
e^{-r/\lambda}\right),
\label{eq26}
\end{equation}

\nn where $\alpha$ is a
dimensionless constant characterizing the strength of the Yukawa force,
and $\lambda$ is its range.  For theories with $n\geq 1$ extra dimensions
$\alpha \sim 1$ and $\lambda \sim R_{n}$, where $R_{n}$ is the size of the
compact dimensions,  Eq.~(\ref{eq26}) holds under the condition
\cite{1}

\begin{equation}
r\gg R_n\sim\frac{1}{M_{Pl}^{(N)}}
\left(\frac{M_{Pl}}{M_{Pl}^{(N)}}\right)^{2/n} \sim
10^{32/n-17}\,\mbox{cm}. \label{eq27}
\end{equation}

\nn For $n=1$ it follows from Eq.~(\ref{eq27}) that $R_1\sim
10^{15}\,$cm which is excluded by solar system tests of Newtonian
gravity \cite{66}. If, however, $n=2$ or $n=3$, the sizes of extra
dimensions are $R_2\sim 1\,$mm or $R_3\sim 5\,$nm, respectively.  While
recent gravity experiments have investigated millimeter distance scales
without finding evidence of new physics, gravity remains poorly
tested at scales $\lesssim 10^{-4}$ m (see Ref.~\cite{66a} for
a review).

For models of non-compact (but warped) extra dimensions \cite{2}
the potential energy takes the form of the Newtonian potential with
a power-law correction

\begin{equation} V(r)=-\frac{Gm_1m_2}{r}\left( 1+
\frac{2}{3k^2r^2}\right),
\label{eq28}
\end{equation}

\nn where $r\gg 1/k$
and $1/k$ is the warping scale.   The correction in Eq.~(\ref{eq28}) can be
generalized to arbitrary inverse powers,

\begin{equation}
V_l(r)=-\frac{Gm_1m_2}{r}\left[ 1+
\alpha_l\left(\frac{r_0}{r}\right)^{l-1}\right],
\label{eq29}
\end{equation}

\nn where $\alpha_l$ is a dimensionless constant, $l$ is a positive
integer,  and $r_0=10^{-15}\,$m.

We note that Yukawa and power-law corrections
given in Eqs.~(\ref{eq26}) and (\ref{eq29}) also arise in ways
unrelated to extra-dimensional physics. For example, the
Yukawa potential describes new forces generated by the
exchange of light bosons of mass $\mu=1/\lambda$, such as
scalar axions, graviphotons, hyperphotons, dilatons, and moduli among others
(see, e.g., \cite{66,66a,67,68,69}). For such forces the interaction
constant $\alpha$ could be much larger than unity. Power-law
corrections, as in Eq.~(\ref{eq29}), arise from the simultaneous
exchange of two photons or two massless scalars ($l=2$ \cite{70}),
two massless pseudoscalars ($l=3$ \cite{71,72}), and from the exchange
of a massless axion or a massless neutrino-antineutrino pair
($l=5$ \cite{72,73}).

The agreement between theory and experiment for our Casimir force
measurements can be used to set new constraints on the Yukawa strength
$\alpha$ as a function of $\lambda$ from
Eq.~(\ref{eq26}). The total force acting between a sphere and a plate due to
the potential described by Eq.~(\ref{eq26}) is obtained by integration over
the volumes of the sphere and the plate, and subsequent differentiation with
respect to
$z$. In fact, the contribution of the Newtonian gravitational force is very
small and can be neglected. To prove this, let us consider a sphere above the
center of an enlarged plate modelled by a disk with a radius
$L_0\gg R=294.3\,\mu$m. (In the present experiment the projection of the
sphere center is displaced by 450\,$\mu$m from one edge of the plate, and
by 50\,$\mu$m from the other). In this case the Newtonian
gravitational force is given by \cite{38}

\begin{equation} F_N\approx -\frac{8}{3}\pi^2 G \rho_{disk}\rho_{sphere}
DR^3\left(1-\frac{D}{2L_0}-\frac{R}{L_0}\right),
\label{eq30}
\end{equation}

\nn where $D=3.5\,\mu$m is the thickness of the plate. To obtain
an upper limit, let us neglect the layered structure of both
test bodies and set $L_0\to\infty$,
$\rho_{disk}=\rho_{Cu}=8.93\times 10^3\,$kg/m${}^3$, and
$\rho_{sphere}=\rho_{Au}=19.28\times 10^3\,$kg/m${}^3$. It then
follows from
Eq.~(\ref{eq30}) that $F_N\approx -3.2\times 10^{-17}\,$N.
This value is four orders of magnitude smaller than the absolute
error of the force measurement in the static regime. Hence, the
contribution of the Newtonian gravitational force can be
neglected at this stage.

For a Yukawa force between a sphere and a plate, the constraints should be
calculated considering the detailed structure of the sphere and plate.
The sphere of density
$\rho_s=4.1\times 10^3\,$kg/m${}^3$ was coated with a layer of Cr of
thickness
$\Delta_{Cr}=1\,$nm with
$\rho_{Cr}=7.19\times 10^3\,$kg/m${}^3$, and  a layer of Au of
thickness $\Delta_{Au}=203\,$nm. The plate of density
$\rho_{Si}=2.33\times 10^3\,$kg/m${}^3$ was coated first with
the same thickness of Cr and then with a layer of Cu of
thickness $\Delta_{Cu}=200\,$nm. Considering that the conditions
$z,\,\lambda\ll R,\,D$ are satisfied, the hypothetical force is
given by ~\cite{34,35}

\begin{equation}
F^{hyp}(z)=-4\pi^2 G\alpha\lambda^3e^{-z/\lambda}R
\left[\rho_{Au}-\left(\rho_{Au}-\rho_{Cr}\right)e^{-\Delta_{Au}/\lambda}
-\left(\rho_{Cr}-\rho_{s}\right)e^{-(\Delta_{Au}+\Delta_{Cr})/\lambda}
\right] \label{eq31}
\end{equation}
\[
\phantom{aaaaaaaa}
\times
\left[\rho_{Cu}-\left(\rho_{Cu}-\rho_{Cr}\right)e^{-\Delta_{Cu}/\lambda}
-\left(\rho_{Cr}-\rho_{Si}\right)e^{-(\Delta_{Cu}+\Delta_{Cr})/\lambda}
\right].
\nonumber
\]

In our experiment, the strongest constraints on the Yukawa hypothetical
interactions are obtained from the dynamic measurement of the
parallel plate pressure (i.e., Casimir force gradient) rather than the
static measurement of the Casimir force.  For this case, the
   Yukawa pressure can be found from Eq.~(\ref{eq31}) by
the using Eq.~(\ref{eq6}), which follows from the proximity
force theorem, Eq.~(\ref{eq5}),

\begin{equation} P^{hyp}(z)=-2\pi G\alpha\lambda^2e^{-z/\lambda}
\left[\rho_{Au}-\left(\rho_{Au}-\rho_{Cr}\right)e^{-\Delta_{Au}/\lambda}
-\left(\rho_{Cr}-\rho_{s}\right)e^{-(\Delta_{Au}+\Delta_{Cr})/\lambda}
\right] \label{eq32} \end{equation}
\[
\phantom{aaaaaaaa}
\times
\left[\rho_{Cu}-\left(\rho_{Cu}-\rho_{Cr}\right)e^{-\Delta_{Cu}/\lambda}
-\left(\rho_{Cr}-\rho_{Si}\right)e^{-(\Delta_{Cu}+\Delta_{Cr})/\lambda}
\right].
\nonumber
\]

\nn Note that the Newtonian gravitational pressure is also below the
sensitivity of the present experiment and can be neglected.

As shown in Ref.~\cite{35}, surface roughness can
significantly influence the magnitude of a hypothetical force in the
nanometer range. To compute the hypothetical pressure taking  account
of roughness, one can use exactly the same method that was applied
in Sec.~IV in the case of the Casimir pressure. The result is

\begin{equation}
P_R^{hyp}(z)=\sum\limits_{i,j=1}^{n}v_iv_jP^{hyp}(z+2H_0-h_i-h_j),
\label{eq33}
\end{equation}

\nn where our notation is that of
Sec.~IV, and $P^{hyp}$ is given by Eq.~(\ref{eq32}).

With these results, we can now  obtain constraints on the hypothetical
Yukawa pressure from the agreement of our measurements of the Casimir
pressure with theory.  According to the results of
Sec.~V, the optimal separation region for obtaining constraints is
$z>310\,$nm. Here the r.m.s. deviation between theory and
experiment, $\sigma_{1100}^{P}=0.34\,$mPa, is somewhat smaller than would be
the case if the complete measurement range were used,  which includes the
shortest separations 260\,nm$\leq z\leq$310\,nm.  (At the shortest
separations the experimental relative error is less than the
theoretical error, but it increases with separation so it eventually
exceeds the theoretical relative error.)  Within the interval
$z>310\,$nm the strongest constraints are obtained from the shortest
separations. We choose
$z_0=320\,$nm and obtain constraints from the inequality

\begin{equation}
\left|P_{R}^{hyp}(z_0)\right|\leq\sigma_{1100}^{P}=
0.34\,\mbox{mPa}.
\label{eq34}
\end{equation}

In Fig.~\ref{constr} constraints on $\alpha$ following from
Eq.~(\ref{eq34}) are plotted for different values of the
interaction range $\lambda$ (curve 1). In the same figure constraints from
previous experiments are also shown. They were obtained  from old
measurements of the Casimir force between dielectrics \cite{34} (curve 2),
from Casimir force measurements by means of a torsion pendulum
\cite{19,35} (curve 3), and by the use of an AFM \cite{23,35,38}
(curve 4). In all cases the region in the ($\alpha,\,\lambda$)
plane above the curve is excluded, and below the curve is
allowed by the experimental results. As can be seen from
Fig.~\ref{constr}, the present experiment leads to the strongest
constraints in a wide interaction range,
56\,nm$\leq\lambda\leq$330\,nm. The largest improvement, by a
factor of 11, is achieved at $\lambda \approx 150\,$nm. We note that the
constraints obtained here almost completely fill in the gap between the
modern constraints obtained by AFM measurements, and those obtained using
a torsion pendulum.  Within this gap the best previous constraints were
obtained from old measurements of the Casimir force between dielectrics
which are not as precise or reliable as those obtained here from the
Casimir pressure measurements between metals using an MTO.

Turning to the power-law-type hypothetical interactions given by
Eq.~(\ref{eq29}), the present experiment does not lead to improved
constraints.   This is explained by the fact that the
metallic coatings used were too thin (and hence, too light) to give
a significant contribution to hypothetical interactions with a longer
interaction range. In fact, the thicker (bulk) matter contributes more
significantly in this case, even though its density is much lower than
for the metal coatings. To obtain stronger constraints on the constants
characterizing new power-law  interactions, thicker metal coatings
and larger interacting bodies are preferable. We anticipate that
future measurements of the Casimir force will employ such samples.

\section{Conclusions and discussion}

Our primary objective in the present paper has been to set new limits on
extra-dimensional models and other physics beyond the Standard Model using
Casimir force measurements between a sphere and a plate separated by
$\sim 0.2$--1.2 $\mu$m.  These experimental results along with a detailed
theoretical analysis lead to new constraints on Yukawa modifications of
Newtonian gravity at short distances, and these are presented in Fig.~14.

Although the constraints on the Yukawa parameter $\alpha$ we obtain
  are $ \sim  10^{13}$, this does not imply that our experiment has
  to be improved by 13 orders of magnitude to detect Newtonian
  gravity.  Rather, the exponential factor
  $e^{-r/\lambda}$ in Eq.(\ref{eq26}) strongly suppresses contributions
  from those parts of the sphere and plate separated by $\gtrsim
  \lambda$.  This ``finite size effect'' \cite{66}, which is not
  relevant for gravity, has a consequence that a Yukawa force between
  the sphere and plate with $\alpha = 1$ is much weaker than gravity.
  The actual gravitational force in our experiment is only $\sim 5$
  orders of magnitude from being detected (using the actual materials
  employed), a gap that may be closed within the forseeable future
  based on the rapid progress various groups have made in the last few
  years.

To carry out our objective, an MTO was used to obtain the first precise
measurements of the Casimir force between dissimilar metals Cu and Au. In the
static regime, the Casimir force between a sphere and a plate was
measured with an absolute error 0.3\,pN at a 95\% confidence
level. This translates into an experimental relative error of the
Casimir force measurements at the  separation 188\,nm of
approximately 0.27\%, i.e., several times smaller than the most
precise previous experiments carried out by means of an AFM
\cite{20,21,22,23}.   To take advantage of the high quality factor of the
MTO,  the Casimir force derivative between the sphere and plate was then
measured dynamically.  This derivative was shown to be
effectively equivalent to the
pressure between two parallel plates composed of the same
materials.   This pressure was determined with a mean absolute
error of 0.6\,mPa at the same confidence level, which leads to an
experimental relative error of approximately 0.26\% at the  separation
260\,nm.
(This compares with a relative error of 15\% quoted by the authors
of Ref.~\cite{26} who directly measured the
Casimir pressure between parallel plates.)

As noted above, the precise calculation of the Casimir force and pressure
between real metals that is needed to extract the effects of new physics
calls for the careful accounting of different corrections
(e.g., due to surface roughness, finite conductivity of the boundary metals,
nonzero temperature, and finite extent of the plate).
The corrections due to roughness
were computed on the basis of the AFM images of the surfaces of the test
bodies. Finite conductivity corrections were calculated  using
tabulated data for the complex index of refraction.  Other corrections were
also estimated (including traditional thermal corrections) and found
to be negligible. The estimated theoretical uncertainty is at the level
of 1\% of the calculated force, i.e., greater than the experimental
uncertainty at the shortest separations used.  Within the
limits of all errors, theory is in good agreement with
experiment.

The level of agreement between theory and experiment was used to
draw important conclusions concerning the influence of thermal effects on
the Casimir force predicted by quantum field theory at nonzero
temperature. Our experimental results lead to a resolution of the
controversy over whether the thermal effects on the Casimir force
between real metals is close to that predicted by
the Matsubara formulation of
quantum field theory for ideal metals, or  is significantly different as
claimed in Refs.~\cite{40,41,42}.  We have shown that the experimental
results contradict  the large, linear in temperature, thermal
corrections predicted in Refs.~\cite{40,41,42}.
Although the sensitivity of the
current experiment is not yet sufficient to detect the
small thermal corrections to the Casimir force and pressure
predicted  for real metals in Refs.~\cite{31,32}, they
are compatible with our experiment.

The good agreement between theory and experiment was then used to set
stronger constraints on hypothetical Yukawa  interactions
predicted by extra-dimensional physics and extensions to the Standard Model.
Existing limits were  strengthened by a factor of up to 11 within
a wide interaction range, from 56\,nm to 330\,nm.  This interaction range
covers the gap between the modern results obtained from Casimir force
measurements using a torsion pendulum and an AFM. The previously known
constraints within this gap  were based on old, and less reliable, Casimir
force measurements between dielectrics.

Our experimental arrangement suggests the need
for additional work to further improve the agreement between
theory and experiment.  We plan  to use smoother and thicker
metal coatings on the surfaces of the test bodies, yielding a  better
characterization of the roughness. This will permit us to carry out
measurements at shorter separations, and to
significantly strengthen the constraints on the predictions of
extra-dimensional physics in a wider interaction range.
The altimate goal of this program to use the iso-electronic effect
\cite{18,38,75,76,77} to suppress the Casimir force so as to improve
our sensitivity to new forces beyond the Standard Model.

\section*{Acknowledgments}

R.S.D. acknowledges financial support from the Petroleum Research
Foundation through ACS-PRF No. 37542--G. The work of E.F. is
supported in part by the U.S. Department of Energy under
Contract No. DE--AC02--76ER071428. G.L.K and V.M.M. are
grateful to the Department of Physics, Purdue University for
kind hospitality. They were partially supported by CNPq 
and Finep (Brazil).


\newpage
\begin{figure}
\caption{\label{fig1}Schematic of the experimental setup showing
its main components, see text. Inset: Resonance curve for the MTO.
Also shown is a Lorentzian fit with $Q \sim $~8000.
See text for further details.}
\end{figure}
\begin{figure}
\caption{\label{circuit}
Schematic of the bridge circuit used to
measure the capacitance. The DC voltages $V_1$ and $V_2$ are used
to correct for small deviations when no interactions are present.
They also linearize the response of the circuit.
Details of the charge amplifier and the part of the circuit to balance
the bridge are omitted for clarity.
Inset: Scanning electron microscopy image of the MTO. }
\end{figure}
\begin{figure}
\caption{\label{deltaV}Dependence of the angular deviation
$\theta$ as a function of  the applied voltage to the sphere. Data
obtained at two different separations $z$ between the metallic
layers are shown. }
\end{figure}
\begin{figure}
\caption{\label{fig2}Electrostatic force $F_{e}$ as a function of
separation $z$ for $\Delta V = V_{Au} -V_0$ = 0.27~V and $\Delta V =
0.35$~V. The fits of the data using Eq. (\ref{eq1}) are also
shown by the solid lines.  See text for further discussion.}
\end{figure}
\begin{figure}
\caption{\label{bandz}Angular displacement of the MTO as a function of
linear displacement when the sphere is moved parallel to the MTO's axis
(top axis, $\bigtriangleup$) and when it is moved perpendicular to
the MTO's axis (bottom axis, $\bigtriangledown$). The data were
acquired at $z = 3~\mu$m with $\Delta V = 0.27$~V. For comparison,
the expected values for an infinite Cu layer using Eq.~(\ref{eq1})
are shown as solid lines. }
\end{figure}
\begin{figure}
\caption{\label{AFM}1$\times 1\,\mu$m$^2$ atomic force
microscopy image of the Cu layer. The topography of the Au layer
on the sphere is similar.  The gray scale to the right of the image gives
the height of  peaks above the bottom of roughness, with
lighter tones corresponding to larger heights.
The set of such images was used to analyze the effect of the surface
roughness on the Casimir force as described in the text. }
\end{figure}
\begin{figure}
\caption{\label{profile} Relief heights $h_i$ versus the fraction
of the total area with heights $h<h_{i+1}$.}
\end{figure}
\begin{figure}
\caption{\label{force}Absolute value of the measured
Casimir force as a
function of separation obtained using the static mode. The value of the
separation between the two metals is determined as discussed in the text.}
\end{figure}
\begin{figure}
\caption{\label{pressure}Absolute value of the parallel plate
Casimir pressure as
a function of separation obtained from the dynamic measurement.
The value of the separation between the two
metals is determined as discussed in the text.}
\end{figure}
\begin{figure}
\caption{\label{deltaF}Difference of the theoretical and experimental
Casimir forces between the sphere and plate versus separation obtained
from the static measurement. }
\end{figure}
\begin{figure}
\caption{\label{deltaP}Difference of the theoretical and experimental
parallel plate Casimir pressures versus separation obtained from
the dynamic measurement. }
\end{figure}
\begin{figure}
\caption{\label{T1}Difference of the theoretical
parallel plate Casimir pressure  as predicted
by Refs.~\cite{40,41} (which incorporates an alternative
thermal correction) and experiment versus separation.
}
\end{figure}
\begin{figure}
\caption{\label{T2}Difference of the theoretical
parallel plate Casimir pressures, as predicted by Ref.~\cite{42}
(which incorporates another alternative
thermal correction) and  experiment  versus separation.
}
\end{figure}
\begin{figure}
\caption{\label{constr}
Constraints on the  Yukawa interaction constant $\alpha$ versus
interaction range $\lambda$. Curve 1 is obtained in this paper,
curve 2 follows from old measurements of the Casimir force
between dielectrics \cite{34}. Curves 3 and 4 are obtained from
the Casimir force measurements between metals by use of
the torsion pendulum \cite{19,35} and by means of an atomic force
microscope \cite{23,38}, respectively. The region in the
($\alpha,\lambda$) plane above each curve is excluded, and below each curve
is allowed.
}
\end{figure}

\begin{thebibliography}{99}
\bibitem{1}
N.~Arkani-Hamed, S.~Dimopoulos, and G.~Dvali,
Phys. Lett. B {\bf 429}, 263 (1998);
Phys. Rev. D {\bf 59}, 086004 (1999).
\bibitem{2}
L.~Randall and R.~Sundrum,
Phys. Rev. Lett. {\bf 83}, 3370 (1999);
{\bf 83}, 4690 (1999).
\bibitem{3}
B.~Abbott {\em et al.},
Phys. Rev. Lett. {\bf 86}, 1156 (2001).
\bibitem{4}
P.~Abreu {\em et al.},
Phys. Lett. B {\bf 485}, 45 (2000).
\bibitem{5}
C.~Adloff {\em et al.},
Phys. Lett. B {\bf 479}, 358 (2000).
\bibitem{6}
C.~Hanhart, D.~R.~Phillips, S.~Reddy, and M.~J.~Savage,
Nucl. Phys. B {\bf 595}, 335 (2001).
\bibitem{7}
S.~Hannestad and G.~G.~Raffelt,
Phys. Rev. Lett. {\bf 87}, 051304 (2001).
\bibitem{8}
S.~Cassisi, V.~Castellani, S.~Degl'Innocenti,
G.~Fiorentini, and R.~Ricci,
Phys. Lett. B {\bf 481}, 323 (2000).
\bibitem{9}
L.~J.~Hall and D.~Smith,
Phys. Rev. D {\bf 60}, 085008 (1999).
\bibitem{10}
M.~Fairbairn,
Phys. Lett. B {\bf 508}, 335 (2001).
\bibitem{11}
S.~Hannestad,
Phys. Rev. D {\bf 64}, 023515 (2001).
\bibitem{12}
A.~Kehagias and K.~Sfetsos,
Phys. Lett. B {\bf 472}, 39 (2000).
\bibitem{13}
E.~G.~Floratos and G.~K.~Leontaris,
Phys. Lett. B {\bf 465}, 95 (1999).
\bibitem{14}
G.~L.~Smith, C.~D.~Hoyle, J.~H.~Gundlach, E.~G.~Adelberger,
B.~R.~Heckel, and H.~E.~Swanson,
Phys. Rev. D {\bf 61}, 022001 (1999).
\bibitem{15}
C.~D.~Hoyle, U.~Schmidt, B.~R.~Heckel, E.~G.~Adelberger,
J.~H.~Gundlach, D.~J.~Kapner, and H.~E.~Swanson,
Phys. Rev. Lett. {\bf 86}, 1418 (2001).
\bibitem{16}
J.~C.~Long, H.~W.~Chan, A.~B.~Churnside, E.~A.~Gulbis,
M.~C.~M.~Varney, and J.~C.~Price,
Nature {\bf 421}, 922 (2003).
\bibitem{17}
J.~Chiaverini, S.~J.~Smullin, A.~A.~Geraci, D.~M.~Weld,
and A.~Kapitulnik,
Phys. Rev. Lett. {\bf 90}, 151101 (2003).
\bibitem{18}
   D.~E.~Krause and E.~Fischbach,
   in {\it Gyroscopes, Clocks, and Interferometers:
   Testing General Relativity in Space}, eds.
C.~L\"{a}mmerzahl, C.W.F.~Everitt, and F.W.~Hehl
(Springer-Verlag, Berlin, 2001).
\bibitem{19}
S.~K.~Lamoreaux, { Phys. Rev. Lett.}
{\bf 78}, 5 (1997); {\bf 81}, 5475(E) (1998).
\bibitem {20}
U.~Mohideen and A.~Roy,
{ Phys. Rev. Lett.}
{\bf 81}, 4549 (1998);
G.~L.~Klimchitskaya, A.~Roy, U.~Mohideen, and V.~M. Mos\-te\-panenko,
{ Phys. Rev. A}
{\bf 60}, 3487 (1999).
\bibitem {21}
A.~Roy, C.-Y.~Lin, and U.~Mohideen,
{ Phys. Rev. D}
{\bf 60}, 111101(R) (1999).
\bibitem {22}
A.~Roy and U.~Mohideen,
{ Phys. Rev. Lett.}
{\bf 82}, 4380 (1999).
\bibitem{23}
B.~W.~Harris, F.~Chen, and U.~Mohideen,
Phys. Rev. A {\bf 62}, 052109 (2000).
\bibitem{24}
T.~Ederth, Phys. Rev. A {\bf 62}, 062104 (2000).
\bibitem{25}
H.~B.~Chan, V.~A.~Aksyuk, R.~N.~Kleiman, D.~J.~Bishop, and F.~Capasso,
Science {\bf 291}, 1941 (2001);
Phys. Rev. Lett. {\bf 87}, 211801 (2001).
\bibitem{26}
G.~Bressi, G.\ Carugno, R.~Onofrio, and G.~Ruoso,
Phys. Rev. Lett. {\bf 88}, 041804 (2002).
\bibitem{27}
F.~Chen, U.~Mohideen, G.~L.~Klimchitskaya, and
V.\ M.\ Mos\-te\-pa\-nen\-ko,
Phys. Rev. Lett. {\bf 88}, 101801 (2002);
Phys. Rev. A {\bf 66}, 032113 (2002).
\bibitem{27a} F. Chen, G. L. Klimchitskaya, U. Mohideen,
and V. M. Mostepanenko, paper in preparation.
\bibitem{28}
A.~Lambrecht and S.~Reynaud,
Eur. Phys. J. D {\bf 8}, 309 (2000).
\bibitem {29}
G.~L.~Klimchitskaya, U.~Mohideen, and V.~M.~Mostepanenko,
{ Phys. Rev. A}
{\bf 61}, 062107 (2000).
\bibitem {30}
V.~B.~Bezerra, G.~L.~Klimchitskaya, and V.~M.~Mostepanenko,
Phys. Rev. A {\bf 62}, 014102 (2000).
\bibitem {31}
C.~Genet, A.~Lambrecht, and S.~Reynaud,
Phys. Rev. A {\bf 62}, 012110 (2000).
\bibitem{32}
M.~Bordag, B.~Geyer, G.~L.~Klimchitskaya,
and V.~M.~Mostepanenko,
{ Phys. Rev. Lett.}  {\bf 85}, 503  (2000);
   {\bf 87}, 259102  (2001).
\bibitem{33}
B.~Geyer, G.~L.~Klimchitskaya,
and V.~M.~Mostepanenko,
Phys. Rev. A {\bf 67}, 062102 (2003).
\bibitem{34}
M.~Bordag, U.~Mohideen, and V.~M.~Mostepanenko,
{ Phys. Rep.} {\bf 353}, 1 (2001).
\bibitem{35}
M.~Bordag, B.~Geyer, G.~L.~Klimchitskaya, and
V.\ M.\ Mos\-te\-pa\-nen\-ko,
{Phys. Rev. D}  {\bf 58}, 075003 (1998);
{\bf 60}, 055004 (1999);
{\bf 62}, 011701(R) (2000).
\bibitem{36}
J.~C.~Long, H.~W.~Chan, and J.~C.~Price,
Nucl. Phys. B {\bf 539}, 23 (1999).
\bibitem{37}
V.~M.~Mostepanenko and M.~Novello,
{Phys. Rev. D}
{\bf 63}, 115003 (2001).
\bibitem{38}
E.~Fischbach, D.~E.~Krause, V.~M.~Mostepanenko, and M.~Novello,
{Phys. Rev. D}
{\bf 64}, 075010 (2001).
\bibitem{39}
R.~S.~Decca, D.~L\'{o}pez, E.~Fischbach, and D.~E.~Krause,
{Phys. Rev. Lett.}
{\bf 91}, 050402 (2003).
\bibitem {51}
E.~M.~Lifshitz,
Zh. Eksp. Teor. Fiz. {\bf 29}, 94 (1956)
[Sov. Phys. JETP  {\bf 2}, 73 (1956)].
\bibitem{40}
M.~Bostr\"{o}m and B.~E.~Sernelius,
Phys. Rev. Lett. {\bf 84}, 4757 (2000).
\bibitem {41}
J.~S.~H{\o}ye, I.~Brevik, J.~B.~Aarseth, and K.~A.~Milton,
Phys. Rev. E {\bf 67}, 056116 (2003).
\bibitem{42}
V.~B.~Svetovoy and M.~V.~Lokhanin,
Phys. Lett. A {\bf 280}, 177 (2001).
\bibitem{43}
C.~A.~Bolle {\em et al.}, Nature {\bf 399}, 43
(1999).
\bibitem{44}
S.~D.~Senturia, {\em Microsystem Design}
(Kluwer Academic Publishers, Boston, 2001).
\bibitem{45}
F.~Ayela {\em et al.}, Rev. Sci. Instrum. {\bf 71},
2211 (2000).
\bibitem{46}
M.~L.~Roukes, Solid-State Sensor and Actuator Workshop
(Hilton Head Island, South Carolina, 2000).
\bibitem{47}
W.~R.~Smythe, {\em Static and Dynamic Electricity}
(McGraw-Hill, New York, 1939).
\bibitem{48}
R.~S.~Decca, H.~D.~Drew, and K.~L.~Empson, Rev.
Sci. Instrum. {\bf 68}, 1291 (1997).
\bibitem{49}
J.~Blocki {\em et al.},  Ann. Phys. (N.Y.) {\bf 105}, 427 (1977).
\bibitem{50}
B.~V.~Derjaguin, I.~I.~Abrikosova, and E.~M.~Lifshitz,
Q. Rev. Chem. Soc. {\bf 10}, 295 (1956).

\bibitem{52}
B.~Geyer, G.~L.~Klimchitskaya,
and V.~M.~Mostepanenko,
Phys. Rev. A {\bf 65}, 062109 (2002).
\bibitem {53}
M.~Schaden and L.~Spruch,
Phys. Rev. A {\bf 58}, 935 (1998).
\bibitem {54}
P.~Johansson and P.~Apell,
Phys. Rev. B {\bf 56}, 4159 (1997).
\bibitem {55}
V.~B.~Bezerra, G.~L.~Klimchitskaya, and V.~M.~Mostepanenko,
Phys. Rev. A {\bf 66}, 062109 (2002).
\bibitem {56}
V.~B.~Bezerra, G.~L.~Klimchitskaya, and C.~Romero,
   { Phys. Rev.} A
{\bf 65}, 012111 (2002).
\bibitem {57}
{\it Handbook of Optical Constants of Solids},
edited by E.D.~Palik (Academic Press, New York, 1998).
\bibitem{58}
T.~Emig, A.~Hanke, R.~Golestanian, and M.~Kardar,
Phys. Rev. Lett. {\bf 87}, 260402 (2001).
\bibitem {59}
V.~B.~Bezerra, G.~L.~Klimchitskaya, and C.~Romero,
Mod. Phys. Lett.  A {\bf 12}, 2613 (1997).
\bibitem{59a}
R. Esquivel, C. Villareal, and W. L. Moch\'{a}n, quant-ph/0306139.
\bibitem{60}
F.~Chen, G.~L.~Klimchitskaya, U.~Mohideen, and
V.\ M.\ Mos\-te\-pa\-nen\-ko,
Phys. Rev. Lett. {\bf 90}, 160404 (2003).
\bibitem{61}
J.~Mehra, Physica (Amsterdam) {\bf 37}, 145 (1967).
\bibitem {62}
L.~S.~Brown and G.~J.~Maclay, Phys. Rev.
{\bf 184}, 127 (1969).
\bibitem{63}
P.~W.~Milonni,
{\it The Quantum Vacuum}
(Academic Press, San Diego, 1994).
\bibitem{64}
V.~M.~Mostepanenko and N.~N.~Trunov,
{\it The Casimir Effect and its Applications}
(Clarendon Press, Oxford, 1997).
\bibitem{65}
K.~A.~Milton, {\it The Casimir Effect}
(World Scientific, Singapore, 2001).
\bibitem{65a}
J. R. Torgerson and S. K. Lamoreaux, quant-ph/0309153.
\bibitem{66}
E.~Fischbach and C.~L.~Talmadge,
{\it The Search for Non-Newtonian Gravity}
(Springer-Verlag, New York, 1999).
\bibitem{66a}
   E.~G.~Adelberger,  B.~R.~Heckel,
and A.~E.~Nelson, hep-ph/0307284;
{Ann. Rev. Nucl. Part. Sci.} {\bf 53} (2003), to appear.
\bibitem{67}
Y.~Fujii, Int. J. Mod. Phys. A {\bf 6}, 3505 (1991).
\bibitem{68}
   E.~G.~Adelberger,  B.~R.~Heckel, C.~W.~Stubbs,
and W.~F.~Rogers,
{Ann. Rev. Nucl. Part. Sci.} {\bf 41}, 269 (1991).
\bibitem{69}
S.~Dimopoulos and G.~F.~Giudice,
{Phys. Lett. B} {\bf 379}, 105 (1996).
\bibitem{70}
J.~Sucher and G.~Feinberg,
in {\it Long-Range Casimir Forces}, eds. F.~S.~Levin
and D.~A.~Micha (Plenum, New York, 1993).
\bibitem{71}
S.~D.~Drell and K.~Huang,
{ Phys. Rev.} {\bf 91}, 1527 (1953).
\bibitem{72}
F.~Ferrer and J.~A.~Grifols,
{ Phys. Rev. D} {\bf 58}, 096006 (1998).
\bibitem{73}
G.~Feinberg and J.~Sucher,
{ Phys. Rev.} {\bf 166}, 1638 (1968);
E.~Fischbach,
Ann. Phys. (N.Y.) {\bf 247}, 213 (1996); S. D. H. Hsu and P. Sikivie, Phys.
Rev. D {\bf 49}, 4951 (1994).
\bibitem{75}
E.\ Fischbach, S.\ W.\ Howell, S.\ Karunatillake,
D.\ E.\ Krause, R.\ Reifenberger, and M.\ West,
Class. Quant. Grav. {\bf 18},  2427 (2001).
\bibitem{76}
D.\ E.\ Krause and E.\ Fischbach, Phys. Rev. Lett. {\bf
  89}, 1904906 (2002).
\bibitem{77}
E.\ Fischbach, D.\ E.\ Krause, R.\ S.\ Decca, and D.\ L\'{o}pez,
Phys. Lett. A., in press.
\end{thebibliography}
\end{document}